# Pore-Spanning Lipid Membrane under Indentation by a Probe Tip: a Molecular Dynamics Simulation Study

Chen-Hsi Huang,[†] Pai-Yi Hsiao,[†,*] Fan-Gang Tseng,[†] Shih-Kang Fan,[‡] Chien-Chung Fu,[§] Rong-Long Pan[#]

[†] *Department of Engineering and System Science, National Tsing Hua University, Taiwan, Republic of China*

[‡] *Department of Materials Science and Engineering, National Chiao Tung University, Taiwan, Republic of China*

[§] *Institute of NanoEngineering and MicroSystems, National Tsing Hua University, Taiwan, Republic of China*

[#] *Department of Life Sciences, National Tsing Hua University, Taiwan, Republic of China*

* Corresponding author. E-mail: pyhsiao@ess.nthu.edu.tw

## Abstract

We study the indentation of a free-standing lipid membrane suspended over a nanopore on a hydrophobic substrate by means of molecular dynamics simulations. We find that in the course of indentation, the membrane bends at the point of contact, and the fringes of the membrane glide downward intermittently along the pore edges and stop gliding when the fringes reach the edge bottoms. The bending continues afterwards, and the large strain eventually induces a phase transition in the membrane, transformed from a bilayered structure to an interdigitated structure. The membrane is finally ruptured when the indentation goes deep enough. Several local physical quantities in the pore regions are calculated, which include the tilt angle of lipid molecules, the nematic order, the included angle and the distance between neighboring lipids. The variations of these quantities reveal many detailed, not-yet-specified local structural transitions of lipid molecules under indentation. The force-indentation curve is also studied and discussed. The results make connection between the microscopic structure and the macroscopic properties, and provide deep insight in the understanding of the stability of a lipid membrane spanning over a nanopore.

# I. Introduction

Lipid molecules are one of the fundamental components in cells. They can aggregate and form cell membranes, which are a prototype of self-assembly systems found in nature.[1,2] A lipid molecule is generally composed of a hydrophilic head group and one or two hydrophobic tails. The assembly of lipid molecules into a two-layered planar structure, with the tails of lipid pointing inward the bilayer and the head groups outward in contact with the watery environments, forms an impenetrable barrier to ions and molecules to maintain unique chemical and physiological environments from the two sides of the bilayer.

During the past decades, the characteristics of lipid bilayers have been wildly studied, such as the features of different phases,[3-6] membrane fusion,[7-9] and phase transitions[10-14]. Temperature usually imposes strong effect on the arrangement of lipid molecules and the structure of lipid bilayers (See, for e.g., Ref. 15 for a review in experimental study and Ref. 16 in simulation study). At high temperature, lipid molecules form a liquid crystalline phase $L_\alpha$ with the tails of molecule disorderedly distributing inside the membrane. This phase is characterized by small membrane thickness, large surface area per lipid, and high mobility of lipid molecule. At low temperature, lipid molecules exhibit a gel phase $L_\beta$, which is a two-layered structure. Depending on the chemical structure of lipids, the gel phase can be categorized to be non-tilted or tilted. In a non-tilted gel phase, the tails of the lipid molecules are perpendicular to the surface of the bilayer, whereas the tails in a tilted gel phase are oblique. Moreover, for some lipid molecules, interdigitation of a gel phase is observed. The interdigitated gel phase, denoted by $L_{\beta I}$, is a phase that the tails of the lipid molecules in the two leaflets of the bilayer interpenetrate, resulting in the extension of the lateral dimension of a membrane.[12] This phase is distinguished by small membrane thickness, very large area per lipid, and high ordering.

Recently, lipid membranes have been integrated into the applications of Micro-electro-mechanical Systems (MEMS).[17-22] Because lipid molecules are the basic materials of cell membranes, they are ideal to serve as a platform in biological research to investigate, for example, the functions of membrane proteins. Conventionally, a lipid membrane in the study is spread over on the surface of a plane substrate. It is called a *solid-supported lipid membrane*.[23-28] Solid-supported lipid membranes are highly stable and homogeneous.[29] However, the presence of the supporting substrate introduces the additional surface effect on the system and the compartment space bellow the membrane does not exist anymore. It hence restricts the possibility to investigate the properties of the lipid membranes in many physiological conditions such as the functions of membrane proteins in transportation of matters, the mechanism of ion flux through a membrane, and so on.[30]

To overcome the problems, free-standing membranes have been proposed in applications.[31-33] In this method, lipid membranes are suspended over pores and hence, the space is divided into, upper and lower, two compartments, similar to the one divided by a real cell membrane. It provides a controllable platform to investigate membrane proteins in a circumstance close to their native environments, without the superfluous surface effect coming from the solid-supported substrate.

Since 2000, different kinds of free-standing lipid membrane systems have been developed for variety of study. Hennesthal and Steinem have applied Scanning Force Microscopy to study the structure of lipid membranes supported by a porous hydrophilic alumina substrate.[31] In their study, the pores were non-pierced cavities on an alumina substrate so that lipid membranes suspended on such pores can only be investigated from the top side. Because the substrate is hydrophilic, the free-standing membranes were entirely supported above the pores instead of spanning over them. It made the applications more restricted in comparison with pore-spanning cases.[34]

The pore-spanning lipid membranes are formed on open hydrophobic pores. Recently, researchers have used them to study embedded membrane proteins. For example, Simon et al. have fabricated arrays of pores on a silicon synthetic layer and formed a bio-mimetic lipid membrane spanning over the pores for the study of indentation by Atomic Force Microscope (AFM).[20] They found that the lipid membranes on small pores are more stable over time than on large pores. Moreover, under mechanical stress, the lipid membranes retain an elastic behavior when suspended on small pores, whereas they are irreversibly deformed when suspended on large pores. Gonçalves et al. have reported a 10% of area expansion of lipid membrane spanning over a 150 nm-diameter pore and a 3-nN yield force when rupture occurs by AFM indentation.[35] They discussed how to use this two-compartment membrane system to study conformational changes of membrane proteins drawn by gradients, cargo transports, and external forces.

Although the free-standing lipid systems have shown great promising in biological research, there are still many difficulties in maintaining the stability of such systems: the suspended lipid membrane is easily broken by the action of an external force or due to small environmental perturbations like temperature fluctuations, substrate vibrations, and so on. Therefore, one crucial problem to apply the platform systems successfully is to increase the stability of the membrane system. To achieve this goal, a thorough knowledge on the mechanical properties of a lipid membrane under bending and stretching is necessary, especially at the level of molecular scale.

A series of early experimental research regarding the mechanical properties of lipid membranes has been conducted since 1981.[36-41] Evans et al. have used a micropipette to study the mechanical properties of a giant vesicle under tension. They found that an abrupt rupture happens when the tension reaches a critical value, and the critical value increases significantly as the loading rate of the tension is high.[37,41] Following the experimental studies, simulations have also been performed to study the topics.[42-44] Neder et al. have employed Monte Carlo simulations to study the conformational transitions of lipid membranes under surface tensions at varies temperatures in bulk solutions.[44] They found that tension can induce interdigitation of a lipid bilayer. This discovery is important in understanding the microscopic properties of lipid membrane which is generally hard to obtain from experiments. Moreover, the mechanical properties of lipid membrane have also been studied by probe indentation. Steinem et al. have systematically studied the indentation of lipid membranes suspended on pores under various conditions by changing the type of lipid molecules, pore size, AFM tips, and so on.[31,34,45,46] They found that the force-indentation curve is governed by the lateral tension (called the prestress effect) when a membrane is suspended on a hydrophobic substrate, which leads to a linear behavior. On the other hand, the force-indentation curve is dominated by bending and stretching stresses when it is suspended on a hydrophilic substrate, which may contribute to a non-linear behavior.

Despite many efforts, the mechanism which causes the rupture of a suspended lipid membrane under tension is still not clear at current stage. The lack of information in the internal structure of the membrane systems leads many open questions. For example, how does the rim of a nanopore interact with the lipid molecules when the membrane spans on it?

Does the internal structure of the membrane, in addition to the external structure, change with indentation by an AFM tip? An insight from the molecular perspective is important in revealing the mechanism and answering the questions.

In this study, we employ non-equilibrium molecular dynamics (MD) simulation to investigate the phenomena of lipid membranes under indentation. The research topics include both the transition of local arrangement of lipid molecules and the study of the membrane morphology. Several structural and mechanical properties are studied. Recently, Chandross et al. have performed molecular dynamics simulations to study nanotribological properties of a self-assembled monolayer (SAMs) on amorphous silica during an indentation process.[47,48] They investigated the dependence of the normal and the friction forces on several factors including the tip size, chain length, and binding interaction. Wallace and Sansom have performed steered MD simulations to study a block of carbon nanotube (CNT) penetrating through a lipid bilayer in bulk solutions.[49] Their study provided useful information concerning the insertion of a hard tubule through a lipid membrane. However, in their work, the effect due to the supporting substrates and the possible rupture mechanism are not discussed. To our best knowledge, this is the first time that the mechanical properties of a lipid membrane suspended on a nanopore is studied by simulations and tested by an indenting probe.

The remaining of the article is organized as follow. In the Sec. II, the simulation model and setup are described. The results are presented and discussed in Sec. III. We first present test runs to verify the validity of the setting of our simulation model (in Sec. III-A). We then present the shape and the overview of a lipid membrane under indentation (in Sec. III-B). The conformational transition of lipid membrane and other related internal properties are studied in Sec. III-C. The force-indentation curves are calculated in Section III-D. The conclusions are given in Sec. IV.

## II. Simulation model and setup

Our system consists of lipid molecules, solvent molecules, and a hydrophobic thin supporting substrate. A pore of nanometer size is opened through the substrate and the lipid molecules are spread across the pore, from the top and bottom sides of the substrate. Due to the hydrophobic interaction, the lipid molecules self-assemble into a bilayered structure, spanning and suspending over the pore. Outside the pore region, the two lipid monolayers sandwich the substrate. The mechanical properties of this pore-spanning lipid bilayer are then studied by indentation with a tip, from the top at the middle of the pore, to imitate a real experimental probing, for example, by AFM.

To simulate such a system by all-atom model simulations, in which the details of the composed atoms are included and every condition of experiments is respected, is nearly impossible because the model contains an enormous number of atoms, which generally requires lots of computing resources and exceeds the power of most of today's computers to handle it. To overcome this difficulty, a coarse-grained approach is used in this study, which groups certain atoms into new interaction units. Moreover, a quasi-2 dimensional (2D) method is employed, in which only a thin slice of the pore-spanning lipid bilayer is simulated. The thin slice models the lipid system transected through the center of the pore. The thickness of the slice is chosen to give lipid molecules degrees of freedom to move, even in the direction of the thickness. Periodic boundary condition is applied in the x-, y- and z-directions to span the modeling system to a 3-dimensional case. The above two approaches significantly reduce the number of interaction sites of the system and the simulations become feasible by today's computing power. The detailed setup of the system is described in the following.

**Lipid molecule:** In the coarse-grained model, neighboring atoms in a lipid molecule are grouped into new units, represented by beads, in a way which maintains the fundamental structure of a lipid, as shown in Fig. 1.

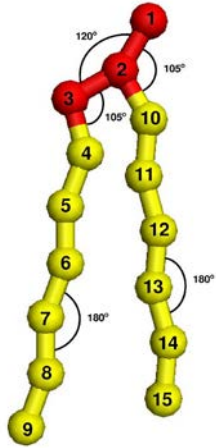

**Figure 1.** A schematic diagram of lipid model. The red beads represent the head group of lipid. The yellow beads represent lipid tails. The values of $\theta_0$ are indicated in the diagram.

A lipid molecule is modeled as a bead-spring, double-tailed molecule. The head of the lipid molecule is composed of 3 hydrophilic beads and each tail of the lipid is composed of 6 hydrophobic beads. All pairs of the beads interact through truncated and shifted Lennard-Jones (LJ) potentials:

$$U_{LJ}^{shifted}(r) = \begin{cases} u_{LJ}(r) - u_{LJ}(r_c) &, r \leq r_c \\ 0 &, r > r_c \end{cases} \quad (1)$$

where $r$ is the distance between two beads and $u_{LJ}(r) = 4\varepsilon\left[(\sigma/r)^{12} - (\sigma/r)^{6}\right]$. The LJ parameters $\sigma$ and $\varepsilon$ denote the diameter of bead and the strength of interaction, respectively. We assume that $\sigma$ and $\varepsilon$ are identical for all pairs of beads. For tail-tail bead interactions, the cutoff $r_c$ is chosen to be $2.5\sigma$, which includes the attractive interaction of the LJ potential. On the contrast, for head-head and head-tail pairs of interaction, $r_c$ is set to $\sqrt[6]{2}\sigma$; in this case, the LJ potential is purely repulsive. This setup models the amphiphilicity of a lipid molecule.

The reason why we do not include the attractive interaction in the head-head pair interaction is to avoid the unphysical wrinkling occurred on the surface of the lipid membrane due to the use of the "phantom solvent model" described below.[11] The beads in the lipid molecule are connected and form a chemical structure shown in Fig. 1. The bead-bead connection is modeled by the finitely extensible nonlinear elastic potential:

$$U_{bond}(b) = -\frac{1}{2}kb_{max}^2 \ln\left[1-\left(\frac{b}{b_{max}}\right)^2\right] \qquad (2)$$

where $b$ is the bond length, $b_{max} = 1.5\sigma$ is the maximum bond length, and $k = 30\varepsilon/\sigma^2$ is the spring constant. This choice of parameters can avoid unlimited bond extension and crossing. Bond angle potentials are incorporated in the model, through a harmonic form

$$U_{angle}(\theta) = k_a(\theta-\theta_0)^2 \qquad (3)$$

where $\theta$ is the angle between two adjacent bonds and $k_a = 2\varepsilon/\text{rad}^2$ is the bending constant. There are three $\theta_0$ in our lipid model: $\theta_0 = 120°$ for the bead triplet 1-2-3, $\theta_0 = 105°$ for the bead triplets 2-3-4 and 1-2-10, and $\theta_0 = 180°$ for the other triplets in the tails, followed with the bead labeling in Fig. 1. We remark that each bead in our coarse-grained model represents a group of chemical atoms. A sketch of the coarse-graining mapping can be found in Ref. 50. Moreover, the choice of these parameters generates a persistence of about $2\sigma$ for the tail chains, which corresponds to the one in a typical DPPC or DMPC lipid molecule.[10] A similar model has been used to study the properties of bilayer membranes[10] and the fusion between two liposomes[7].

**Solvent molecule:** Solvent environment is simulated by explicit solvent beads. A bead represents one cluster of three water molecules.[10] In order to avoid immaterial problem of solvent structure and reduce the simulation time,[51] the "phantom solvent model" is employed in this study.[6,11] The computation time is largely saved through working with this model. In this model, the solvent beads do not interact with each other, but interact with other beads in the simulation box, including the head and the tail beads of the lipid molecules, the beads which constitute the substrate and the indenting probe, via the purely repulsive LJ potential given in Eq. (1) with $r_c = \sqrt[6]{2}\sigma$. Please notice that the model does not allow momentum transfer between solvent molecules.[44] The hydrodynamic effect is hence not properly considered. Nonetheless, recent studies have shown that some hydrodynamic characteristics, especially the conservation of momentum, are very important in the simulations of membrane structure to produce correctly the undulations of lipid membranes.[16,50] Our solvent model holds well this crucial characteristic. Same models have been successfully used in many simulations to study phase and structural transitions of lipid membrane under tension.[6,11,44] Of course, hydrodynamics could bring in some unexpected effects to the system, which can be justified only when it is implemented appropriately. Recent developments have allowed hydrodynamics to be effectively simulated, for example, by dissipative particle dynamics[52,53] or by multi-particle collision (MPC) dynamics[54,55]. Particularly, the MPC dynamics differ to our simulations only by additional collision steps, after streaming steps, which permit the momentum transfer between solvents. However, these collision steps request more computing efforts. Limited to our resources, we will not use MPC dynamics in this study. Regarding that the momentum conservation is held in our model, we do not expect significant changes in our results if MPC dynamics is implemented.

**Substrate and probe:** The substrate and the probe consist of beads. The beads form simple cubic lattice structure in them, with lattice constant equal to 1σ. In our quasi-2D model, the nanopore is represented by two rectangular blocks of substrate located separately on the left and right hand-sides of the simulation box (See initial configuration in Fig. 2(a)). The distance between the inner edges of the two blocks is the diameter of the pore and is set to 80$\sigma$. The thickness of the substrate is 20$\sigma$ and the width of each block is 30$\sigma$. We consider the case that the substrate surface is hydrophobic. The substrate beads interact with the lipid tail beads through the attractive version of Eq. (1) with $r_c = 2.5\sigma$. They interact repulsively with the other beads. The probe is shaped as a hemispherical body and the radius of the hemisphere is 15$\sigma$. It is used in the study of indentation. The probe beads interact repulsively with other kinds of beads. In a real case, the probe can have adhesion with lipid and solvent molecules.[45,56] There is also the proposal that solvent extrusion can give rise to the extra attraction.[57] The hydrophobic or hydrophilic character of probe has been shown to affect the details of membranes under indentation or intrusion.[45,47,49,56] Nonetheless, in this study, we pay more attention on the pure response of the lipid membranes to the indenting force. The attraction force with the probe is ignored. Therefore, no adhesion between the probe and lipid membranes and solvent is considered.

A summary of the interactions between different types of beads used in this simulation can be found in Table 1.

**Table 1.** Summary of interactions between lipid head beads, lipid tail beads, solvent beads, substrate beads, and probe beads. The notation R means a repulsive interaction given by Eq. (1) with $r_c = \sqrt[6]{2}\sigma$. The notation A means an attractive interaction given by Eq. (1) with $r_c = 2.5\sigma$. X means no interaction.

| Bead-Bead | Head | Tail | Solvent | Substrate | Probe |
|---|---|---|---|---|---|
| Head | R | R | R | R | R |
| Tail | R | A | R | A | R |
| Solvent | R | R | X | R | R |
| Substrate | R | A | R | X | R |
| Probe | R | R | R | R | X |

We place 570 lipid molecules and 27530 solvent beads in the simulation box of size $140\sigma \times 6\sigma \times 50\sigma$. The smallest side of the simulation box (in the y-direction) allows 3 lipid molecules to fit in, side by side. Because the periodic boundary condition is applied, our quasi-2D model simulates a system with infinite replicas. Therefore, three dimensional effect has been largely taken into account. The density of solvent corresponds to the water density in the ambient condition. The simulations are performed in NVT ensemble using Nosé-Hoover thermostat.[58,59] The equations of motion are solved by Verlet integrator using simulation package LAMMPS.[60] The integration time step $\Delta t$ is $0.005\tau$, where $\tau = \sigma\sqrt{m/\varepsilon}$ is our time unit and $m$ is the mass of a bead. The temperature T is set to $0.9\varepsilon/k_B$, except in Sec. III-A in which T is varied to study the phase behavior, where $k_B$ is the Boltzmann constant. Initially, the lipid molecules are positioned into two layers with the lipid tails pointing inward, jointing at the middle of the pore as shown in Fig. 2(a). After about $10^7$ MD steps of equilibration, the lipid molecules form a bilayered structure, suspending across the pore. We will show in Sec. III-A that the lipid membrane is in a gel phase at this temperature. In the study of indentation, the probe is placed in the solvent region where the solvent beads have been removed previously. The initial position of the probe is at the center, $10\sigma$ above the middle plane of the pore (see Fig. 2(b)).

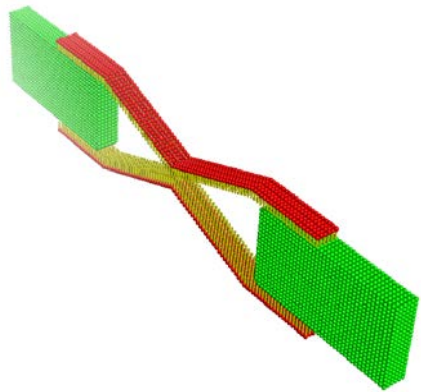

(a)

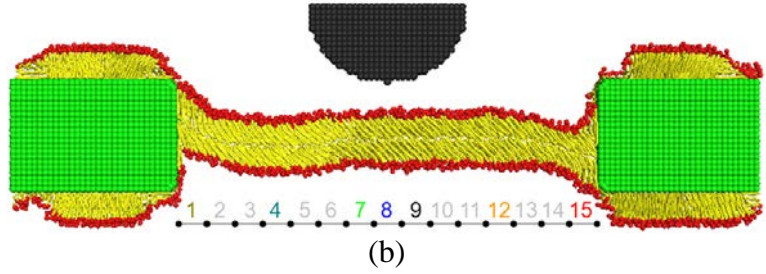
(b)

**Figure 2.** (a) Initial configuration. (b) Snapshots of the system before indentation. The red and the yellow beads represent the head groups and the tails of lipid molecules, respectively. The green beads constitute the pore substrate and the black ones the hemisphere probe. The solvent beads in the picture are not shown for clarity. The lipid molecules are initially placed into two layers which finally sandwich the substrate across the pore when the system is equilibrated. The pore is divided into 15 regions, as indicated in (b). In Sec. III, we will study the local structure of lipid membrane in these regions.

After equilibration, the probe moves downward at a speed of 0.001 $\sigma/\tau$. The indentation process runs for $10^7$ MD steps. At the final stage of the process, the probe pierces through the membrane. 5 independent runs were performed in this study for average and check of the consistence. The simulations were run on a home-made PC cluster. A typical run of indentation takes about 400 hours of CPU time. In our model, $\sigma$ and $\tau$ correspond to a real length of about 0.46 nm and a time of 2.1 ps, respectively. Therefore, we simulate a membrane suspended on a nanopore of diameter of 37nm. Here the speed of indentation is very large, several orders of magnitude faster than a typical speed used in experiments. This is a typical problem encountered in steered MD simulations, due to the limitation of today's computing power.[47,49] Despite the fast indentation rate used here, the response of the lipid membrane in our study exhibits an elastic behavior, similar to what have been observed in experiments with a low rate.[34] Therefore, our simulations can qualitatively capture the properties of a real system. For the first simulation study of the indentation of the system, we concentrate on the generic behavior and phenomena revealed at molecular level by the indentation.[61] To shorten the notation, in the following text, all the physical quantities will be reported in a reduced unit system in which the length unit is $\sigma$ and the energy unit is $\varepsilon$. For example, the strength of force is reported in unit of $\varepsilon/\sigma$.

## III. Results and discussions

### A. Phase transition of lipid membrane in bulk solutions

Before entering into the main topics of study in the indentation of a suspended lipid membrane, we study firstly the phase behavior of lipid membranes in bulk solutions in order to justify the setting of our model.

The systems were run in isothermal-isobaric ensemble using Nosé-Hoover barostat with the pressure setting up at $1.0$.[62] The initial size of the simulation box was chosen to be $100 \times 6 \times 40$. Since the dimension in y-direction is significantly small, compared to the ones in the x- and z-directions, the system can be regarded as a quasi-2D system. Lipid molecules were initially positioned at the lattice points of a two-layered hexagonal lattice with the head group of the lipids pointing outward. Two systems consisting of different amounts of lipid molecules, 270 and 330 lipids, were simulated. 9000 solvent beads were added into the two systems. Each system was run at several temperature points, ranging from 0.8 to 1.15. It took about $10^7$ simulation steps to get the systems equilibrated.

After equilibration, we collected data for the following $2 \times 10^7$ steps to calculate membrane thickness, area per lipid, and nematic order parameter to study the structure and the phase of the lipid bilayer. Membrane thickness is defined to be the average vertical distance between the surfaces of the first tail beads in the upper and the lower leaflets of the bilayer. The first tail beads are labeled by 4 and 10 shown in Fig. 1. Area per lipid is calculated by dividing the twice of the cross section of the system in the xy-plane by the number of lipid molecules. The tensor of nematic order in a studied region of space is

calculated by

$$S_{\alpha\beta} = \frac{1}{2N_l} \sum_{n=1}^{N_l} (3\ell_{n\alpha}\ell_{n\beta} - \delta_{\alpha\beta})$$

where the sum runs over all the $N_l$ lipid molecules in the region, $\ell_{n\alpha}$ and $\ell_{n\beta}$ are, respectively, the $\alpha$ and $\beta$ components of the direction vector $\vec{\ell}_n$ of the $n$th lipid molecule, $\alpha$ and $\beta$ stand for the vector components ($x$, $y$, or $z$) in Cartesian coordinates, and $\delta_{\alpha\beta}$ is the Kronecker delta.[63] The direction vector $\vec{\ell}_n$ is calculated by averaging the two tail vectors $\vec{r_9 r_4}$ and $\vec{r_{15} r_{10}}$, in Fig. 1, of a lipid molecule. The tensor was calculated from simulations, and three eigenvalues of the tensor were computed. The largest eigenvalue is called the *nematic order parameter*.[11] The biggest possible value of the nematic order parameter is 1, which results from a perfect arrangement of the direction vectors pointing parallel to each other. The value 0 derives from a completely randomization of the vectors.[63] The results for the membrane thickness $D$, the area per lipid $A$, and the nematic order parameter $\eta$ versus temperature $T$ are shown in Fig. 3, panel (a), (b) and (c), respectively. Red curves are the results obtained from the system of 330 lipid molecules, while the blue ones illustrate the system of 270 lipids.

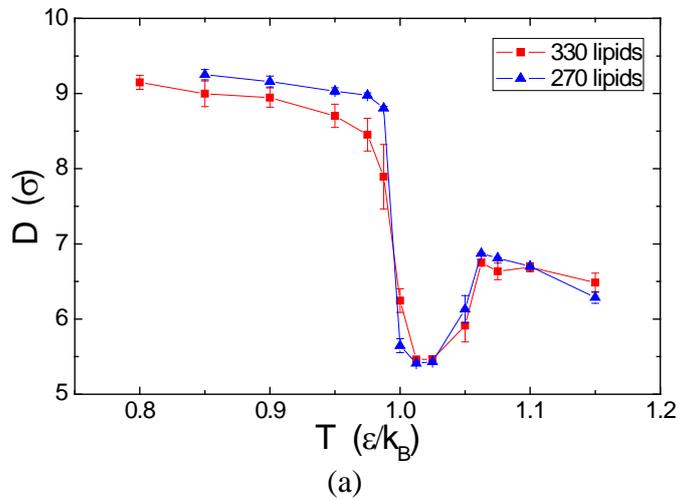

(a)

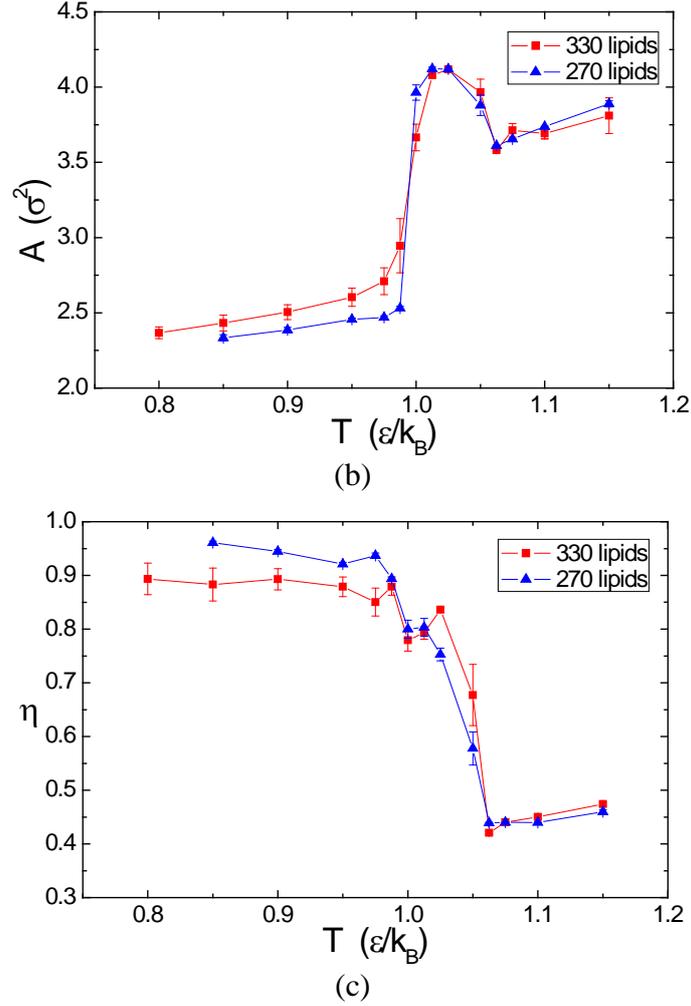

**Figure 3.** (a) membrane thickness D, (b) area per lipid A, and (c) nematic order parameter $\eta$, as a function of temperature T. The results for the system with 330 lipid molecules and the one with 270 lipid molecules are plotted in red and in blue curves, respectively.

We can see that the lipid membrane system exhibits, in turn, three phases with increasing temperature. When the temperature is smaller than 1.0, the lipid membrane shows bilayer structure with the thickness around $9\sigma$ and the area per lipid around $2.5\sigma^2$. The nematic order parameter is about 0.9, which is close to 1, showing a good alignment of the lipid molecules. The system is hence in a gel phase. With *T* rising over 1.0, a drastic decrease happens in the thickness whereas the nematic order parameter is still high. It indicates the formation of the interdigitated gel phase, characterized by small thickness of about $5.5\sigma$ and large area per lipid of $4.25\sigma^2$. We mention that between gel phase and interdigitated gel phase, a two-phase coexistence was observed. In this region, the membranes are partially in the gel phase and partially in the interdigitated phase. When $T$ is larger than 1.05, the state of liquid crystalline phase was observed. In this phase, the lipid membrane has the thickness of about $6.5\sigma$ and the area per lipid of $3.75\sigma^2$. The thickness is thicker than the one in the interdigitated gel phase, but the area per lipid is smaller. More importantly, a small nematic order around 0.4 is observed, showing that the tails of the lipid molecules are arranged in a quite random way inside the bilayer structure. The area per lipid obtained in our simulations corresponds to a real value of $53\text{Å}^2$, $90\text{Å}^2$ and $79\text{Å}^2$ in the gel, interdigitated gel, and liquid crystalline phase, respectively. The results are in agreement with experimental findings[64,65] and simulations.[10]

In order to give readers a clear picture, we present in Fig. 4 the snapshots of membrane in the gel phase, the two-phase coexistence, the interdigitated gel phase, and the liquid crystalline phase. We can see clearly that a bilayered membrane changes its structure to the interdigitated gel state and then, the disordered liquid state as temperature increases.

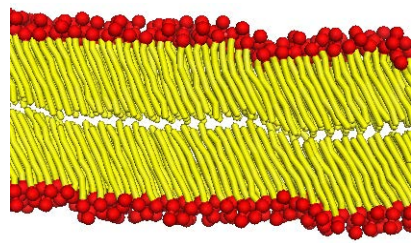

(a)

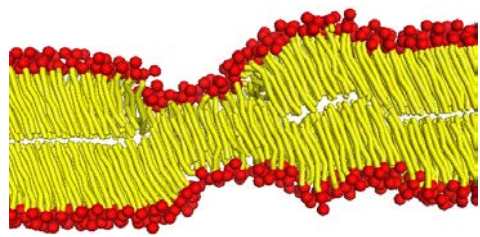

(b)

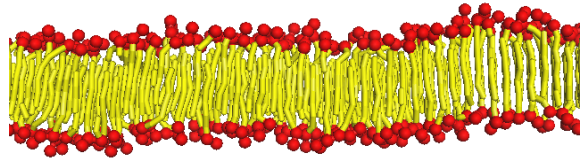

(c)

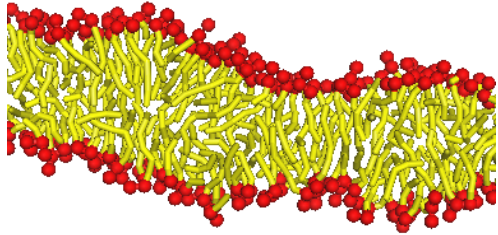

(d)

**Figure 4.** Snapshots of membrane in (a) the gel phase (T=0.9), (b) two-phase coexistence (T=0.975), (c) the interdigitated gel phase (T=1.025), and (d) the liquid crystalline phase (T=1.1). The representation of the picture is the same as described in the caption of Fig. 2.

In the gel phase, we also observed that the lipid molecules have tilting-angle preference. This is essentially due to the large and asymmetric head group in our lipid model. The lipid molecules, commonly used in experiments such as DPPC, DMPC, do show this tilting tendency in the gel phase at low temperature. The tilted gel phase plays an important role in the determination of the bending mechanism of a suspended lipid bilayer, as we will show in Sec. III-C.1. The tilt angle obtained here is 26±7° to the membrane normal, which agrees fairly with experiments[66] and simulations[12]. We have tested the tilting angle dependence upon the balance between head-bead repulsion and tail-bead attraction by varying the head-bead repulsion strength from $0.1\varepsilon$ to $5\varepsilon$, and found that the tilt angle increases not significantly, from 23° to 27°. It is because the structure is mainly determined by the balance between the size of the lipid head group and the tails. To be a general model, we simply set the strength ratio for the head-bead repulsion and tail-bead attraction to be 1:1. This setup is also similar to the setup used by other simulation groups.[7,10,11,44]

The phase behaviors shown here are consistent with those found in real experiments[64,65] and other simulations[6]. The model also vitally captures the titling-angle feature of lipid molecules in membranes. These characteristics support the validity of our simulation model. However, we have to stress that the aim of this work is to provide a general understanding of the behavior and mechanism of a membrane system under indentation through a simple, coarse-grained model. To pursue quantitative agreement for every physical quantity is not our goal and also impossible.

## B. Global structure of suspended lipid membrane under indentation

We now go back to the main topics, studying the indentation of a lipid membrane suspended across a nanopore. The temperature $T$ is fixed at 0.9 at which the system is in a gel phase. Under the indentation, the shape of the membrane deforms and varies with time. To understand this deformation, we calculated the z-coordinate of the middle line of the membrane sheet at every point in the pore. This quantity is denoted by $z_m(x)$ where $x$ is the position across the pore region. The results at different simulation moments are presented in Fig. 5. Each curve displays the membrane shape at a specific time.

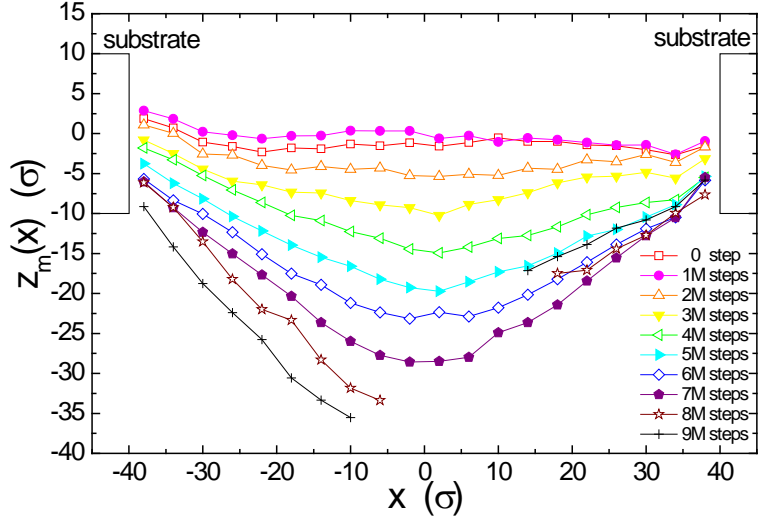

**Figure 5.** $z_m(x)$ at different simulation moments for each 1M timesteps.

We observed that the shape function $z_m(x)$ is not exactly a horizontal line before the indenting probe contacts the membrane. It inclines toward one side of the pore. This inclination occurs because the constituted lipid molecules are not perpendicular to the bilayer when the bilayer is formed. In this situation, the forces exerted on the lipid molecules by the attraction of the pore edges from the two sides shear the membrane in the vertical direction, resulting in an inclined membrane as shown in the snapshot of Fig. 2(b). With time passing, the probe descends. Once it touches the membrane, the membrane starts to deform. The process takes place in two ways: (1) bending, (2) downward-gliding. The bending occurs at the middle of the membrane where the probe contacts with it. The gliding occurs at the two edges of the pore where the membrane moves downward with time. The simulation timestep 0.8M to 5.1M (M stands for one million) in Fig. 5 is referred to this process. The gliding stops when $z_m(x)$ approaches the edge bottom of the pore. Because of the inclination, the right fringe of the membrane stops gliding at timestep 4.0M, earlier than the left fringe.

We plot in Fig. 6 the time evolutions of $z_m(x)$ at the left and the right fringes of the membrane. We saw that the two curves exhibit a step-by-step descending behavior, particularly for the left fringe. It shows that the gliding of the membrane happens intermittently, dropping down suddenly at specific points. Although the probe keeps pushing downward the membrane, the fringes of the lipid membrane remain fixed at a position until the stored stress is large enough to overcome a threshold value, which is the maximum frictional force for the lipid molecules to glide on the pore edge. Mey et al. have reported the similar stair-like descending behavior in the force indentation curves in their experiments.[34]

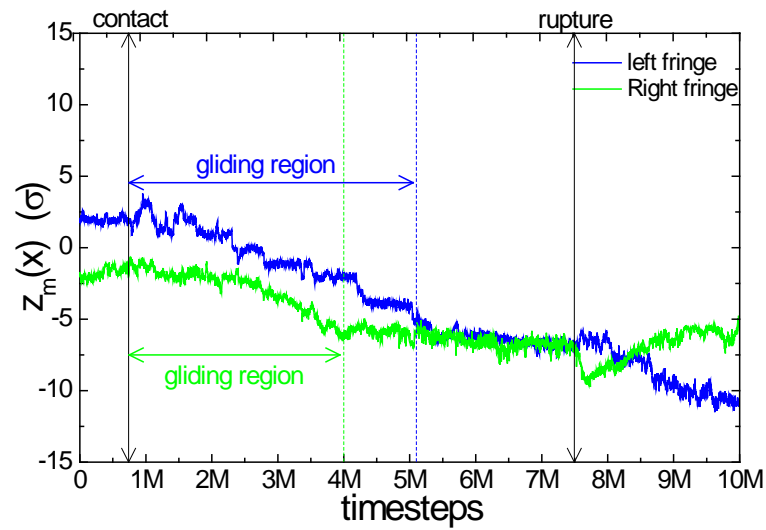

**Figure 6.** Time evolution of $z_m(x)$ at left and right fringes of the membrane, plotted in blue and green colors, respectively. The right fringe stops gliding around timestep 4.0M while the left one around 5.1M.

When the two sides of the membrane both reach the bottoms of the pore edges, the membrane stops gliding. The only way to proceed is the bending. (See timestep 5.1M to 7.5M). The strain increases with time and the membrane becomes more and more arched. When the indentation becomes deep enough, an abrupt rupture takes place because the strain exceeds the mechanical threshold of the membrane, and the membrane is cut into two pieces (Refer to time step 7.5M). We present snapshots of time evolution of the membrane in Fig. 7. These pictures, from top to bottom, show three stages of shape variation for the suspended membrane under indentation: (I) gliding and bending, (II) purely bending, and (III) rupture.

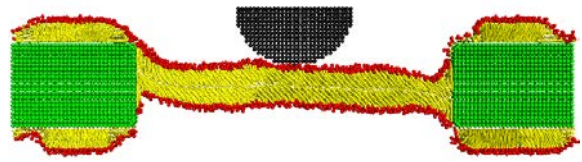

(a)

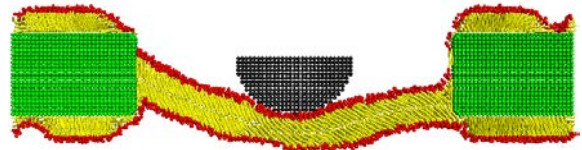

(b)

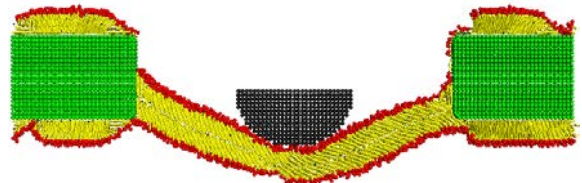

(c)

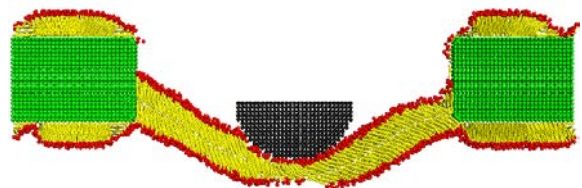

(d)

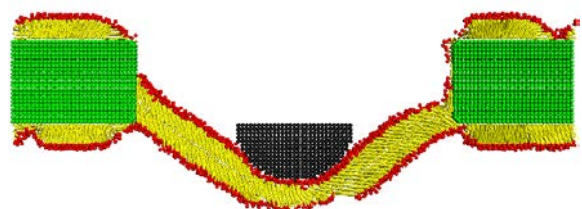

(e)

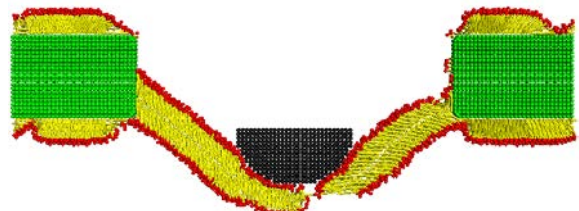

(f)

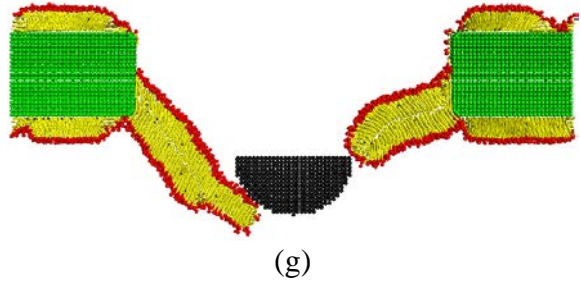

(g)

**Figure 7.** Snapshots of the membrane system under indentation at several timesteps: (a) 1M, (b) 4M, (c) 5.5M, (d) 6M, (e) 7M, (f) 7.5M, and (g) 9M. The representation of the picture is the same as described in the caption of Fig. 2.

## C. Internal structure of suspended membrane under indentation

After having studied the global structural change of the lipid membrane under indentation, we go now inside the membrane and investigate its internal structural change and the local organization of the constituted lipid molecules. We divided the pore into 15 regions along the $x$-direction, as indicated in Fig. 2(b). Region 1 and Region 15 are the two fringe regions where the membrane is in contact with the pore edges; Region 8 is the middle region where the probe indents on the membrane. We calculate the following structural quantities.

**C.1 Tilt angle of lipid molecule**

The tilt angle $\theta$ of a lipid molecule is defined to be the acute angle between the lipid direction vector $\vec{\ell}$ and the $z$-axis. The $z$-axis is perpendicular to the pore-substrate plane in this study. The moving direction of the indenting probe is downward, following $-\hat{z}$ direction. How the tilt angle changes with indentation in different pore regions is plotted in Fig. 8.

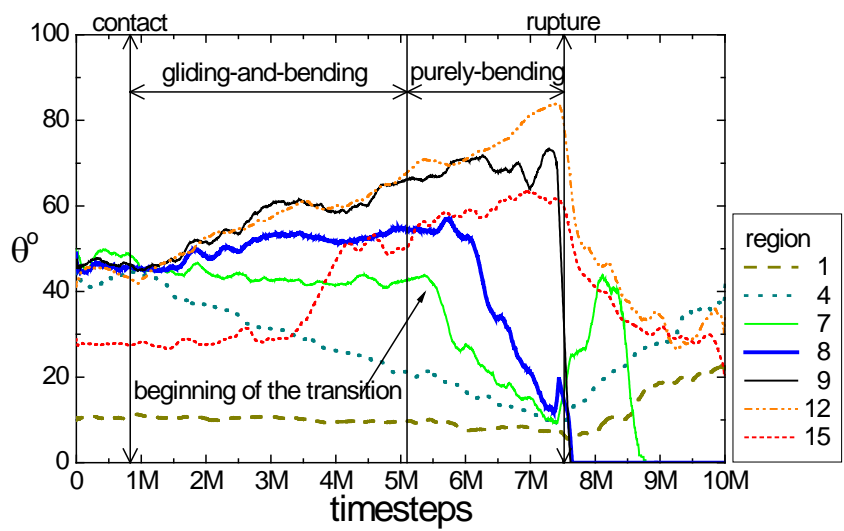

**Figure 8.** Time evolution of the tilt angle θ of lipid molecules in different pore regions.

We observed that before the indenting probe touches the membrane, the tilt angle of the lipid molecules is about 45° in the internal regions of membrane (from Region 2 to 14) with an angle fluctuation of about 10°. This tilt angle is slightly larger than the one in bulk solutions (~35°) obtained by experiments[66] and simulations[12]. It is known that our membrane is subjected to a lateral tension due to the attractive force of the substrate, which pulls outward the membrane from the two pore edges. It renders a lying-down effect on the lipid molecules, in contrast to the membrane in a bulk solution. The lipid molecules thus acquire a larger tilt angle to balance this extra tension. On the other hand, in Region 1 and Region 15, θ is 10° and 27°, respectively. The hydrophobicity of substrate attracts the lipid tails, which renders the lipid molecules lying parallel to the wall edges of the pore, resulting in a small value of θ. In the course of indentation, an asymmetric behavior of angular variation with time was observed: the tilt angle decreases against the indentation for the left regions of the lipid membrane (from Region 1 to Region 6), but increases for the right regions (from Region 9 to Region 15). The physics can be understood as follows. Due to the unique tilting direction of the lipid molecules, the left part of the suspended membrane suffers a shearing strain by the probe indentation, which shears the membrane along the lipid direction. The lipid molecules thus rotate clockwise, leading to a decrease of θ. In contrast, the right part of the membrane suffers a bending strain, which bends the membrane in a way similar to extend an accordion's bellows left-downward. The lipid molecules in this part hence rotate counterclockwise, and θ increases. In the middle regions of the membrane (Region 7 and 8), the lipid molecules tend to maintain their tilt angles against the indentation. Nonetheless, the angle shows an abrupt change, starting at some point in the purely-bending stage. We will explain this phenomenon later.

θ in the two fringe regions, Region 1 and Region 15, does not show significant change at the stage of gliding-and-bending. Membrane gliding can decrease the stress of the indentation and hence, reduce the deformation. This effect is important in the maintenance of the stability of the lipid membrane, particularly at the early stage of indentation, to avoid rupture due to large stress and strain.

We remark that many membrane systems comprise lipid molecules with tilting angle.[16,66-68] Our simulations predict an interesting asymmetric local structural change upon indentation for such the systems. There are also membranes which are made of non-tilted lipid molecules.[15,16] For these systems, the response to the indentation is expected to be different. It is worth to be investigated in the future.

**C.2 Nematic order parameter**

In order to understand the degree of local arrangement of lipid molecules in the suspended membrane, we calculated the nematic order parameter η by the method described in Sec. III-A and studied its variation with time. The results at different pore regions are presented Fig. 9.

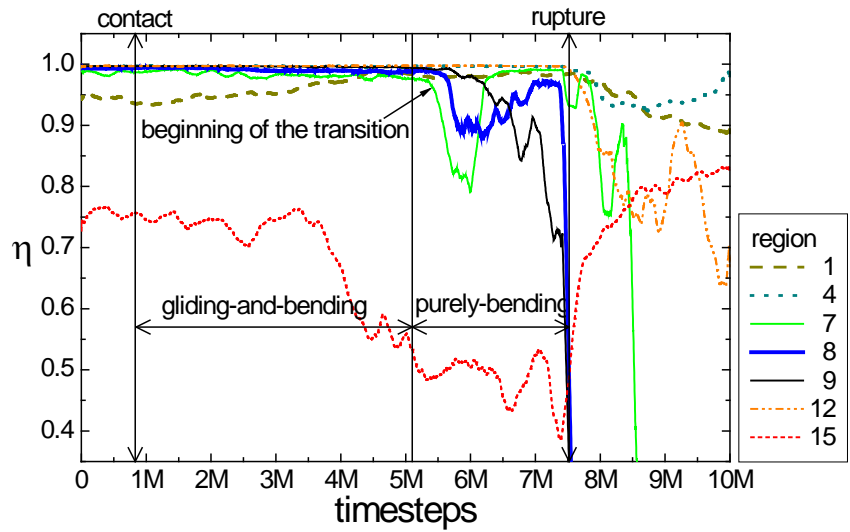

**Figure 9.** Time evolution of the nematic order parameter η in different pore regions.

We observed that at the gliding-and-bending stage, η takes a value very close to 1 for all the regions except the two fringe regions. It suggests a regular arrangement in the interior of the membrane, in which the lipid molecules tend to align parallel to each other. For the lipids in the fringe regions, the molecules also interact with the hydrophobic walls of the pore, which perturbs the alignment. Consequently, the arrangement is less ordered. Please notice that η in the right fringe, Region 15, shows a decreasing behavior, starting at the time step ~4.0M. This behavior occurs because the membrane fringe stops gliding when arriving at the bottom corner of the pore and some of the lipid molecules traverse into the lower surface of the substrate, which results in a decrease in the ordering. On the other hand, the left fringe of the membrane (Region 1) stops gliding at a later moment (at the time step 5.1M). The tilting direction of the lipid molecules on this side is more parallel to the pore wall, which prevents the traversing of the lipid molecules into the lower substrate surface. Therefore, the ordering stays high.

At the purely-bending stage of indentation, a drastic change was observed: the values of η in Regions 7, 8 and 9 drop suddenly at some moment and return back to the value 1 after a while, exhibiting a valley-like curve. This drastic change suggests the occurrence of a structural transition inside the membrane. During the indentation process, the lipid membrane suffers an increasing bending strain by the indenting probe. Gaps eventually occur in the middle of the membrane between the lipid molecules along the x-direction, and grow with time. Since the lipid molecules have more room to fluctuate, the ordering in these regions decreases. When the gap size is large enough, the interpenetration of the lipid molecules in the upper leaflet into the lower leaflet becomes allowed, promoted by the inter-molecular interaction. The bilayered structure is re-assembled into an interdigitated, monolayer structure. This interdigitated layer is a highly ordered structure. Consequently, the order parameter grows back to 1. The structural transition from a bilayered structure into an interdigitated monolayer extends the lateral dimension of the membrane significantly, which decreases indispensably the stress of indentation and prolongs the lifetime of the membrane against the final rupture. A similar stress-induced phase transition has been reported in simulation studies.[44] Neder et al. found that lipid membranes in liquid crystalline phase and in ripple phase can partially or completely transit into an interdigitated phase under surface tension, but not in gel phase. In our simulations, the probe indents the lipid membrane constantly. The stress can be significantly large, which leads the occurrence of such phase transition even from a gel phase.

In our study, the purely-bending stage can be further divided into three phases, depending on the internal structure of membrane: (1) a bilayer phase, (2) a transient phase, and (3) an interdigitated-layer phase. These three phases occur at different time points of indentation for different regions of the membrane. We can see from the snapshots in Fig. 7 how these three phases take place one after the others. We observed that the phase transitions start at Region 7 and propagate to the neighbor regions 8 and 9. The three regions are the most stressed regions on the membrane because they suffer the direct indentation of the probe there. The propagation of the phase transitions can be seen in Fig. 9 where the value of η in Regions 7, 8 and 9 exhibits, in turn, a valley-like curve. When the bilayered structure transforms locally into the interdigitated-layer structure, the lipid molecules inside line up again in a parallel fashion. Moreover, the direction of the lipid molecules gradually become more perpendicular to the pore as having been recorded in Fig. 8 where the tilt angle θ decreases after the occurrence of the phase transition in the regions. The calculated value of θ show that at the moment before the rupture of the membrane, the lipid molecules in these central regions are basically normal to the surface.

The phase transition of lipid membrane under indentation has been observed in experiments. Simon et al found that a part of the sample of lipid membranes in a nanopore in their experiments were irreversibly deformed, which can be attributed to the phase transition found in our simulations.[20] Besides, another experiments performed by Mey et al. also revealed some features of phase transition.[34] It will be discussed in Sec. III-D.

**C.3 Included angle between neighboring lipid molecules**

The occurrence of the phase transition from a bilayered structure to an interdigitated one can be identified by studying the included angle between neighboring lipids. Two kinds of included angle were calculated. The first one, denoted by ϕ, is obtained by calculating the *acute* angle between neighboring lipid molecules without considering their vector directions.

The second one, denoted by φ, is calculated, on the contrast, by considering the vector direction of the neighboring molecules. Therefore, φ takes a value ranging from 0° and 180°. Since the lipid molecules in the upper as well as in the lower leaflet of a bilayered membrane are locally parallel to each other, the value of φ is expected to be 0°. For an interdigitated membrane, the neighboring lipids are anti-parallel and hence, the value of φ should be 180°. However, we expect that ϕ takes a value of 0° for both of the two cases. The mean values of ϕ and φ as a function of time in different pore regions are presented in Fig. 10, panel (a) and (b), respectively.

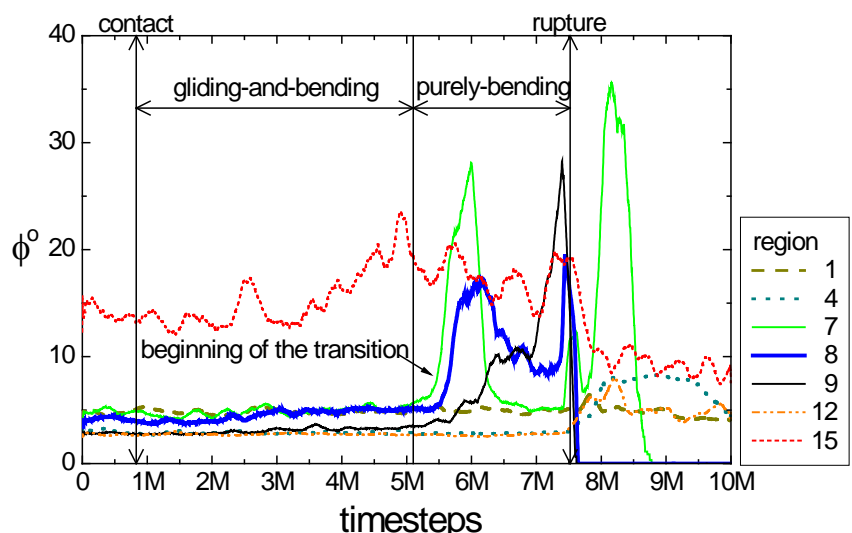

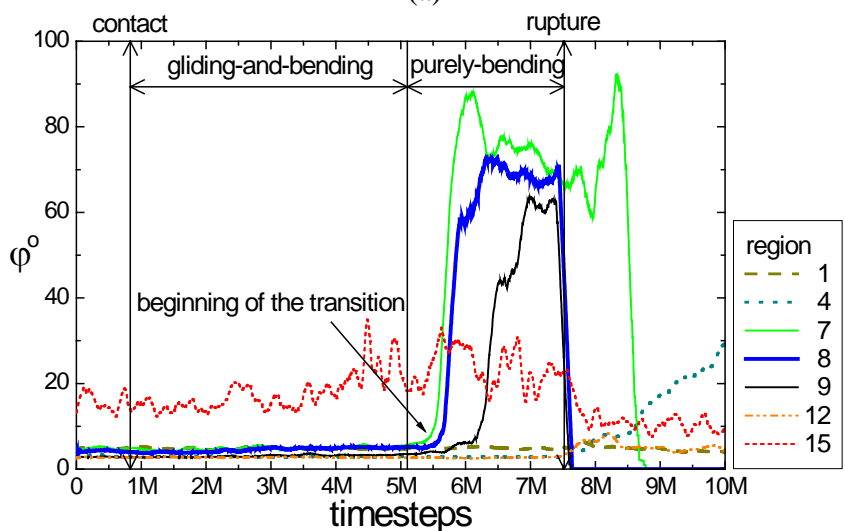

**Figure 10.** Time evolution of the included angles, (a) ϕ and (b) φ, between neighboring lipid

molecules in different pore regions.

We can see that $\phi$ is a constant at the gliding-and-bending stage of indentation in each pore region. It begins to increase when the indentation enters into the purely-bending stage. This increase starts in the middle regions of the membrane and propagates gradually toward the edge regions. As we have seen in the previous subsection, the strain due to the indentation can become very large and eventually triggers a structural transition inside the membrane to reduce the strain. What we can see in Fig. 10(a) is that $\phi$ in the middle regions shows an abrupt increasing and decreasing at the purely-bending stage. These abrupt changes correspond to the same moments when the tilt angle $\theta$ (in Fig. 8) starts to decrease.

The value of $\varphi$ in Regions 7, 8 and 9 also rises up suddenly at the same time (See Fig. 10(b)). It then maintains at a value between about 70° and 90° while $\phi$ drops back to a small value of about 5°. This information, together with the one obtained in the calculation of the ordering parameter ($\eta \approx 1$) in Fig. 9, shows the internal structural change from an ordered (bilayer) structure into another ordered (interdigital) one under the indentation. One question has to be answered: why the value of $\varphi$ is not 180° after the transition as what we expected. The reason is that in our quasi-2D system, the suspended membrane suffers the tensional strain only from the x-direction (across the pore), but not from the y-direction. As the probe goes down, the indenting tension causes the increase of the separation distance of the neighboring lipid molecules in the x-direction. The increase of distance eventually allows the docking of the upper leaflet of the bilayered membrane into the lower leaflets. The interdigitated structure is hence formed along the x-direction, but not the y-direction. Therefore, a lipid molecule has only about half of its neighbors parallel to it and half of it anti-parallel. To confirm this picture, we calculated the percentage $\rho$ of the neighboring lipids in parallel to a surrounded lipid molecule. The results are shown in Fig. 11.

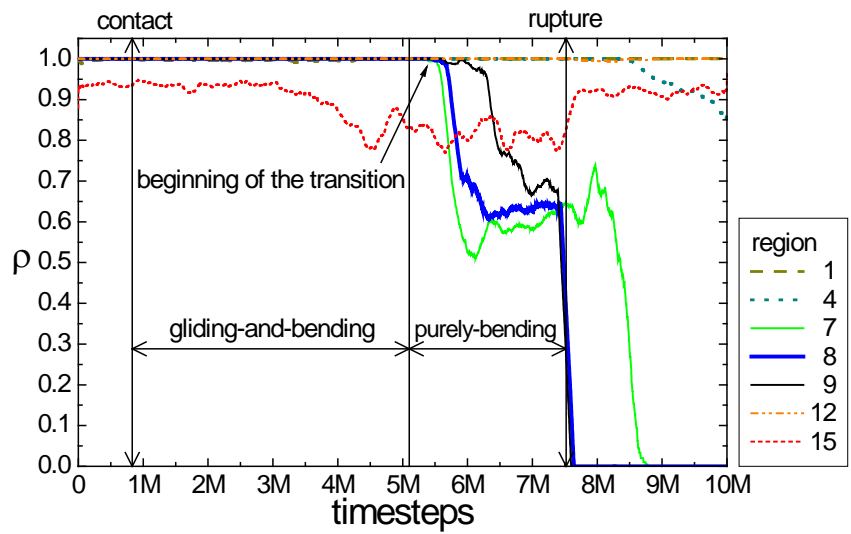

**Figure 11.** Time evolution of the same-direction percentage ρ in different pore regions.

We found that ρ is a constant, very close to 100% at beginning. It agrees with the results of small value of $\phi$ and $\varphi$. At the moment when the structural phase transition takes place (where $\phi$ and $\varphi$ drastically increase), ρ decreases quickly, down to a value of about 60% and maintains basically at this value before rupture of the membrane. The result clearly shows that the neighboring lipid molecules are partially parallel and partially anti-parallel to the surrounded lipid, as what we depicted. Since only about 40% of the neighboring lipids are anti-parallel, the included angle $\varphi$ is estimated to take a value of $0°\times 60\% + 180°\times 40\% = 72°$, which is consistent with the result shown in Fig. 10(b).

It is worth to notice that in a real 3D system, a suspended lipid membrane can suffer lateral tension from the pore rim in both x- and y-directions. Consequently, the interdigitating of the lipid molecules could happen in the two directions upon indentation and propagate outward from the indenting center. The above studied quantities are thus expected to show a pattern depending on the tilting direction of the lipid molecules. This interesting phenomenon can be studied only when a true 3D model of membranes spanning over pores is used. It deserves a further investigation in the future.

**C4. Distance between neighboring lipids**

During the indentation process, the membrane is bent at the middle and the tensional force causes the extension of the lateral dimension of the membrane. As discussed in the previous subsections, this extension enlarges the separation distance between neighboring lipids, which plays an important role in determination of the internal structure of the membrane. To understand this effect quantitatively, we investigated here the separation distance between neighboring lipids. The distance is defined to be the mean distance between the centers of mass (CM) of two neighboring lipids. It can be split into two components. One is the normal distance, which is the distance normal to the direction of the lipid molecules. The other component is the shear displacement, which is the CM distance projected on the lipid direction. The former component describes the normal strain inside the membrane and the latter evaluates the shear strain. The results for the normal distance and the shear displacement are presented in Fig. 12, panel (a) and (b), respectively.

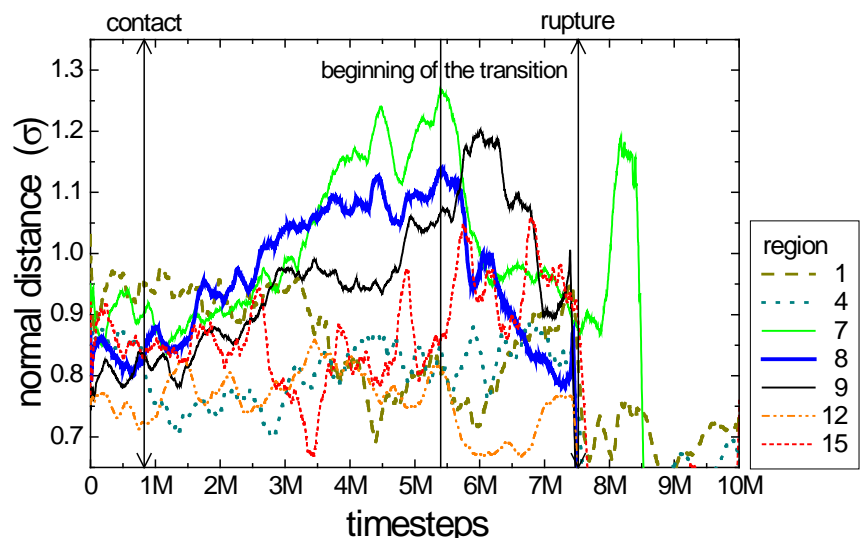

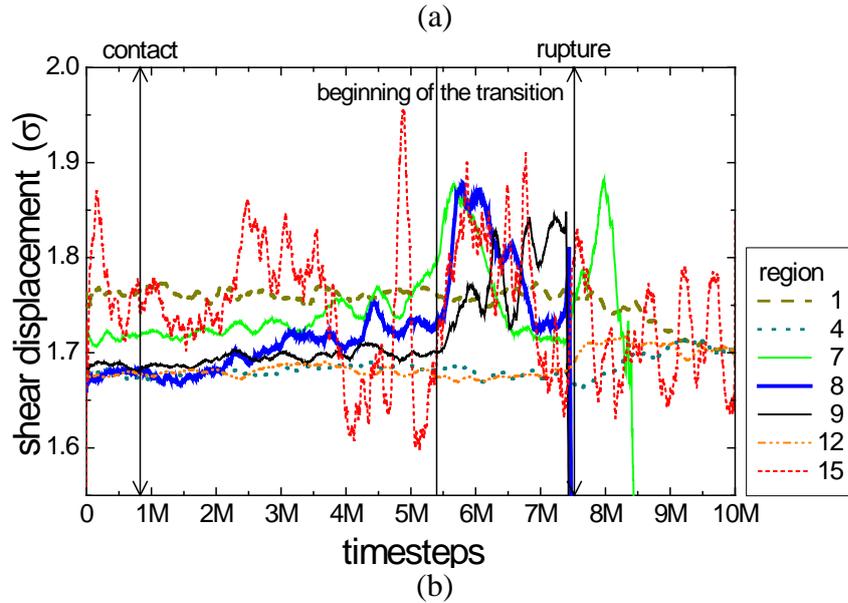

(a)

(b)

**Figure 12.** Time evolution of (a) the normal distance and (b) the shear displacement in different pore regions.

We can see that the normal distance in the middle regions 7, 8, 9 increases in the course of indentation, up to the time step ~5.4M. This increase creates *gaps* between lipid molecules in the bottom leaflet of the membrane. When the gaps are large enough, the lipid molecules in the top leaflet of the membrane penetrate into the gaps in the bottom leaflet, forming the interdigitated-layer structure. Once the new structure is formed, the normal distance decreases and returns back to a value similar to the one in the bilayered structure. Therefore, the curve of the normal distance exhibits a peak structure which marks the occurrence of the phase transition.

The calculation for the shear displacement in different pore regions show that the shear displacement in the middle regions also reveals a peak structure associated to the phase transition. After the formation of the interdigitated phase, the shear displacement turns to decrease. This behavior can be associated with the decrease of the tilt angle of the lipid molecules in the well-ordered interdigitated layer shown in Fig. 8 and Fig. 9. We also observed that Region 7 is the region which reaches firstly the largest shear displacement and normal distance. Therefore, the phase transition takes place firstly in this region. Since the shear displacement is relatively small compared to the normal distance, it is the normal distance which determines the occurrence of the transition.

## D. Force-indentation curve

Finally, we investigated the mechanical property of the system. The relationship between the force and the displacement of the indentation probe was studied and the apparent spring constant was calculated from the slope of the force-indentation curve. The issue is concerned in many applications.

The force of the probe was computed by the reacting force of the lipid molecules acting on the probe. Only the z-component of the force takes effect because the indenting probe moves downward along the z-direction. The indentation depth is the descending distance of the probe counting from its position before indentation. The results, the force versus the indentation depth, are presented in Fig. 13.

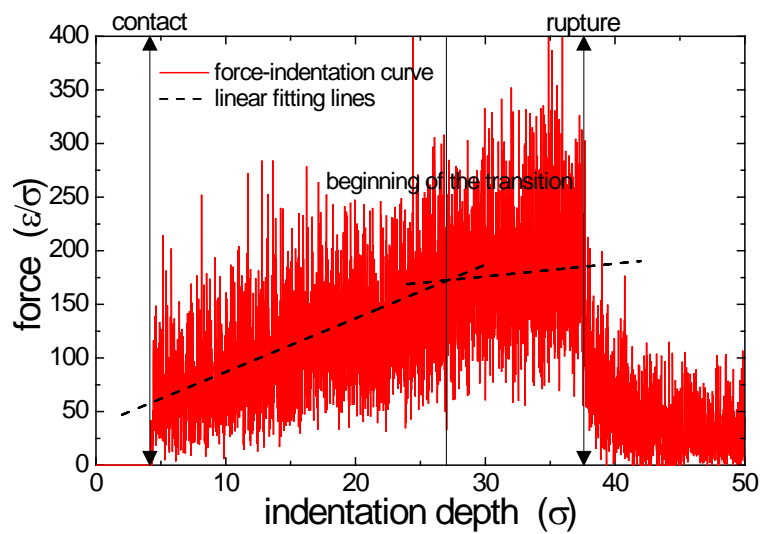

**Figure 13.** Force-indentation curve. Two linear regimes are identified before rupture of the membrane. The dashed lines show the slopes of the linear fitting in the two regimes.

We observed that the force increases with the indentation depth after the probe contacts with the lipid membrane. Two linear regimes were identified. The boundary of the two regimes is located at the moment when the phase transition to an interdigitated state takes place, at the time step 5.4M, or equivalently, at the indentation depth 27$\sigma$. The first regime has a slope much larger than the second one. The two slopes averaged from 5 independent runs yield a value equal to 4.1±0.5 $\varepsilon/\sigma^2$ and 0.2±0.7 $\varepsilon/\sigma^2$, respectively, which correspond to an apparent spring constant of 0.081±0.009 N/m and of 0.003±0.013 N/m, each. The force suddenly drops down when the membrane is ruptured.

The elastic response of membrane upon indentation generally comprises three contributions: bending, lateral tension, and stretching.[69] Bending takes a minor contribution on the apparent spring constant in the range of 0.01mN/m, and the latter two are responsible for a large extent in experiments. Lateral tension gives rise a linear response, which is due to the adhesion and friction of the pore substrate to the suspended membrane. Stretching occurs when the indenting force is strong, and is dominated by a cubic law on the force-indentation curve. Our results show a predominated linear dependence, particularly in the first regime. It is thus the lateral tension which dominates the behavior, as suggested by Mey et al.[34] Moreover, the value of apparent spring constant depends much on the hydrophobic/hydrophilic character of pore substrate. On hydrophilic pores, the spring constant is small and ranges in 0.2~2 mN/m, whereas on hydrophobic pores, it is large and the value stays in a range of 20~40 mN/m.[34,46,56] Our simulations show the same order of magnitude for the force constant.[34,45] In experiments, a non-linear effect on the force-indentation curve has been observed, but usually shows up in a weak contribution.[56]

This effect can be explained by a generalized Canham-Helfrish model, which has been derived recently.[70] However, due to the fluctuations, the nonlinearity cannot be clearly seen in our study. Therefore, following the suggestion of Mey et al.,[34] we pre-assumingly fit the regime by a linear equation to get the apparent spring constant.

Generally, a force-indentation curve can reveal the information about the membrane structural transition in the indentation process. The first linear regime shows that the lipid membrane is in an elastic response against the indentation. The second regime shows a reduction of slope in the force-indentation curve, which indicates a change of internal structure in the lipid membrane. As we have seen in Sec. III-C, a local phase transition from a bilayerd structure to an interdigitated one does occur at this moment. It involves a rearrangement of lipid molecules, which increases the lateral dimension of the membrane upon indentation. This dimensional extension prevents the increase of stress inside the membrane; therefore, the apparent spring constant largely decreases. We note that in experimental and theoretical analysis of force indentation, people usually do not consider any phase transition that changes the internal structure of membrane. Our molecular simulations clearly show this possibility. The effect of the phase transition can be captured in the force-indentation curve by a reduction of the force constant. The data in the second regime of our figure does support this idea. Experiments have already observed a plateau or sawtooth-like behavior in force-indentation curves.[20,34] The behavior could be attributed to the consequence of an internal structural transition. The information obtained here provides a new direction of thinking and analysis of the membrane properties in the future.

## IV. Conclusions

We have established a quasi-2D coarse-grained model for pore-spanning lipid membranes to investigate the mechanical properties of the system under indentation by a semispherical probe. The model has been verified by studying the membrane thickness, the area per lipid, and the nematic order parameter at different temperatures. Three phases of the lipid membrane have been identified: the gel phase, the interdigitated gel phase, and the liquid crystalline phase. In the study of indentation, we have calculated the shape function of the lipid membrane. According to the variation of the shape function, the indentation process can be categorized into three stages: (1) gliding-and-bending stage, (2) purely-bending stage, and (3) rupture stage. In the first stage, the membrane glides downward intermittently and stops gliding when the membrane fringes reach the bottom of the pore edges. Membrane bending occurs at the middle of the membrane due to the indentation of the probe. It takes place in the first and the second stages, and ends up when the membrane is ruptured. We have further investigated the structural properties of the internal lipid molecules. We found that the lipid molecules on the left side of the indentation point rotate clockwise during the indentation, while it rotates counterclockwise on the right side. This behavior occurred because the lipid molecules tilt to the left side in this study; therefore, the left part of the membrane suffers mainly a shearing strain in the course of indentation, whereas the right part suffers a bending strain. Moreover, the local nematic order parameter in the central regions of the membrane exhibits on a valley-like pattern in the purely-bending stage. It indicates the happening of internal structural transition. This transition was further confirmed by calculating two kinds of included angle between neighboring lipid molecules. Lipids in the central regions transformed from a bilayered structure to an interdigitated gel structure. The transition progressively propagates to the neighboring regions. Consequently, the purely-bending stage is divided into three phases: bilayer phase, transient phase, and interdigitated-layer phase. The study of the CM distance between neighboring lipid molecules showed that the lipid molecules get apart from each other under indentation. By separating the distance into the normal distance and the shear displacement, we found that the normal distance determines the occurrence of the transition. An interdigitated phase is formed when the normal distance becomes large enough to allow docking between the lipid molecules in the upper and lower leaflets of the bilayered membrane. Finally, the force-indentation curve has been studied. The results showed two linear regimes. In the first regime, the lipid membrane behaves elastically. The apparent spring constant calculated from the line slope shows consistency with experiments. In the second regime, the internal phase transition takes place. The extension of membrane dimension upon indentation reduces the lateral stress and hence, the apparent spring constant is small. Unlike metal or ceramic, lipid membranes are flexible, soft and diffusible. These characteristics give them the possibility to adapt themselves against external actions through structural transformation or reassembly. The study performed here opens a window for the details of lipid membranes under indentation at molecular level, and provides valuable information in understanding the stability of pore-spanning lipid membranes.

In perspective, there are several topics to be investigated. For example, a true 3D model which includes correct hydrodynamic effect should be used in the future to clarify some points in our quasi-2D model. How will the structural transition of membrane be formed under a probe indentation with respect to the tilting direction of lipid molecules? Will the strain be propagated outward from the indenting center, respecting some kind of geometrical symmetry, said, ellipsoidal? How do the hydrophobic or hydrophilic characters of probes and pore substrates affect the response of membrane? What is the role of tilting of lipid molecules on the indentation study? And a theoretical analysis should be developed to explain the possible impact of structural transition on force-indentation curves.

## Acknowledgements

This material is based upon the work supported by the National Science Council, the Republic of China under Contract Nos. 97-2627-M-007-005, 98-2627-M-007-004, and 99-2627-M-007-004.